%% file: qcdlp.tex
\documentstyle[preprint,epsfig,eqsecnum,aps,floats,tighten]{revtex}

\def\Missing#1#2{{\mbox{$#1\kern-0.57em\raise0.19ex\hbox{/}_{#2}$}}}

\def\vMissing#1#2{\ifmmode
            \vec{#1}\kern-0.57em\raise.19ex\hbox{/}_{#2}
         \else
            {{\mbox{$\vec{#1}\kern-0.57em\raise.19ex\hbox{/}_{#2}$}}}
         \fi}
\def\lsim{\mathrel{\rlap{\lower4pt\hbox{\hskip1pt$\sim$}}
    \raise1pt\hbox{$<$}}}        
\def\gsim{\mathrel{\rlap{\lower4pt\hbox{\hskip1pt$\sim$}}
    \raise1pt\hbox{$>$}}}

\def\D0{D\O }

\newcommand{\Et}{\mbox{$E_{\rm T}$}}

\newcommand{\inb}{\mbox{nb$^{-1}$}}
\newcommand{\rs}{\mbox{$\sqrt{\rm s}$}}

\def\simge{\mathrel{\rlap{\raise 0.53ex \hbox{$>$}}%
{\lower 0.53ex \hbox{$\sim$}}}}
\def\simle{\mathrel{\rlap{\raise 0.53ex \hbox{$<$}}%
{\lower 0.53ex \hbox{$\sim$}}}}

\def\inb{nb$^{-1}$}                     

\def\ETmiss{\mbox{${\hbox{$E$\kern-0.5em\lower-.1ex\hbox{/}\kern+0.15em}}_T$ }}

\newcommand{\pbarp}{\mbox{$p\overline{p}$}}

\def\err#1#2#3 {{\it Erratum} {\bf#1},{\ #2} (19#3)}
\def\ib#1#2#3 {{\it ibid.} {\bf#1},{\ #2} (19#3)}
\def\nc#1#2#3 {Nuovo Cim. {\bf#1} ,#2(19#3)}
\def\nim#1#2#3 {Nucl. Instr. Meth. {\bf#1},{\ #2} (19#3)}
\def\np#1#2#3 {Nucl. Phys. {\bf#1},{\ #2} (19#3)}
\def\pl#1#2#3 {Phys. Lett. {\bf#1},{\ #2} (19#3)}
\def\prev#1#2#3 {Phys. Rev. {\bf#1},{\ #2} (19#3)}
\def\prl#1#2#3 {Phys. Rev. Lett. {\bf#1},{\ #2} (19#3)}
\def\rmp#1#2#3 {Rev. Mod. Phys. {\bf#1},{\ #2} (19#3)}
\def\zp#1#2#3 {Zeit. Phys. {\bf#1},{\ #2} (19#3)}     
\def\inb{nb$^{-1}$}                     

\begin{document}

%
%
\title{
Multiple Jet Production at Low Transverse Energies in \pbarp \ Collisions 
at {\rs} = 1.8 TeV}
\author{\centerline{The D\O\ Collaboration
  \thanks{Submitted to the {\it International Europhysics Conference
        on High Energy Physics},
	\hfill\break
	July 12-18, 2001, Budapest, Hungary,
        \hfill\break 
	and  {\it XX International Symposium on Lepton and Photon Interactions at High Energies}
	\hfill\break
        July 23 -- 28, 2001, Rome, Italy. 
	 }}}
\address{
\centerline{Fermi National Accelerator Laboratory, Batavia, Illinois 60510}
}
%
%
\date{\today}

\maketitle

%
%
\begin{abstract}
We present data on multiple jet production for transverse energies 
greater than 20 GeV in {\pbarp} collisions at {\rs} = 1.8 TeV. 
QCD calculations in the parton shower approximation ({\sc pythia}) and 
in the next-to-leading order approximation ({\sc jetrad}) show 
discrepancies with data for three and four-jet production. This 
disagreement is especially apparent in multiple jet angular and 
transverse momentum distributions. 
\end{abstract}

\newpage 

\begin{center}
\input{list_of_authors_1_june_2001.tex}
\end{center}

\normalsize

\vfill\eject

The study of jet production at high transverse energy was 
one of the main goals of the 1993--1995 run of the Fermilab Tevatron 
collider, and the results have been compared with leading-order QCD 
predictions by both the CDF\cite{ref_1} and 
D\O \cite{ref_2} collaborations. These high-{\Et} data, where {\Et} is 
the transverse energy of the jet were described 
satisfactorily by complete tree-level leading order $2\rightarrow N$ 
QCD calculations\cite{ref_3} and by the {\sc herwig} parton shower 
Monte Carlo\cite{ref_4}. In this paper, we describe studies of the 
complementary kinematic region of $Q^2/\hat s \ll 1 $, where $Q^2$ is 
the square of the momentum transfer between partons, which we set equal 
to $E_{T}^2$, and $\hat s$ is the square of center of mass energy 
in the rest frame of the collision. Here the 
{\sc BFKL}\cite{ref_5} description of jet production differs 
significantly from that of the high-{\Et} {\sc DGLAP}\cite{ref_6} 
kinematic domain of $Q^2 \sim \hat s$. Measurement of jet production 
in this kinematic region can provide information on the evolution of 
higher-order jet processes.   

We present results that extend our previous measurements of 
multiple jet production to lower {\Et}. The data were collected with 
the D\O \ detector during 1993--1995 at a proton-antiproton 
center-of-mass energy of 1800 GeV. Jets were 
measured in the liquid-argon calorimeter, which has a segmentation of 
$\Delta \eta \times \Delta \phi =0.1\times 0.1$, where $\eta $ is 
pseudorapidity and $\phi $ is azimuthal angle\cite{ref_7}.
At least one calorimeter trigger tower 
($\Delta \eta \times \Delta \phi =0.2\times 0.2$) with 
${E_{\rm T}} \ge$ 2 GeV was required by the Level-1 trigger, and at 
least one jet with ${E_{\rm T}} \ge $ 12 GeV was required by the 
Level-2 trigger\cite{ref_11_0}. Jets were reconstructed using a fixed 
cone algorithm with radius 
$\Delta {\cal {R}}=\sqrt{\Delta \eta ^2+\Delta \phi ^2}=0.7$ in 
$\eta -\phi $ space. The jet reconstruction threshold was 
8 GeV. If two jets overlapped and the shared 
transverse energy was more than 50\% of the transverse energy of the 
lower-energy jet, the jets were merged; otherwise they were split into 
two jets. The integrated luminosity of this data sample was 
1.96$\pm $0.29 {\inb}. Instantaneous luminosity was restricted to be 
below 3 $\times 10^{30}cm^{-2}s^{-1}$ to minimize multiple {\pbarp} 
interactions.   

To provide events of high quality, online and offline selection 
criteria were used to suppress multiple interactions, cosmic ray 
backgrounds, and spurious jets. Jets were restricted to the 
pseudorapidity interval $|\eta | \le 3$.  

Jet energies have been corrected for calorimeter response, shower 
development, different sources of noise, and contributions from the 
underlying event\cite{ref_8}. These corrections comprise  
the largest source of systematic uncertainty on the jet cross section. 
The typical value of the correction to jet energy is 15 - 30\%, 
with an uncertainty of 2-4\%. In our study, we consider jets with 
${E_{\rm T}} > $ 20~GeV. For an $n$-inclusive jet event, the 
$n$ leading jets must have transverse energy above the threshold value. 
The trigger efficiency is 0.85 for the inclusive ($n$ = 1) jet sample 
for energies near threshold, rising rapidly to unity at larger {\Et}. 
The efficiency is essentially unity for $n > 1$.     

To compare with data, Monte Carlo (MC) events were generated using the 
{\sc pythia} 6.127 \cite{ref_pythia} and {\sc jetrad}\cite{ref_13} 
programs. These generators simulate particle-level jets in the 
parton-shower approximation, and parton-level jets in the 
next-to-leading order approximation, for {\sc pythia} and {\sc jetrad}, 
respectively. The smearing of jet transverse energies was implemented 
using the experimentally determined jet energy resolution\cite{ref_8}, 
which is $\approx $ 20\% at {\Et} = 20 GeV. In {\sc pythia}, jets were 
reconstructed at the particle level using the D\O\ algorithm, and in 
{\sc jetrad}, at the parton level, using the Snowmass algorithm
\cite{ref_11_0}. 
 
\begin{figure}[h]
\vspace {-1.0cm}
\mbox{\hspace*{1.0cm}\epsfig{figure=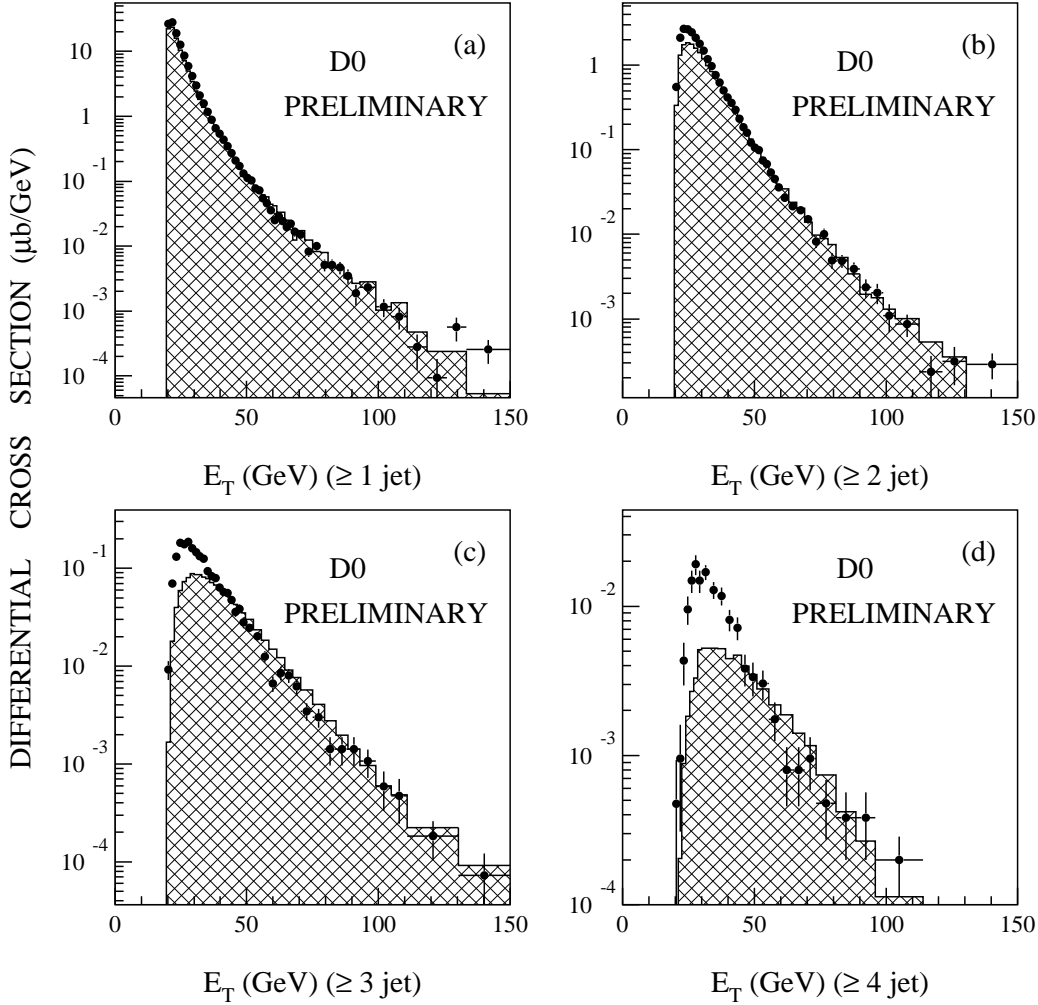,width=15cm}}
\caption{
Distributions in the transverse energy of the leading jet for (a) 
single-inclusive, (b) two-jet inclusive, (c) three-jet inclusive and 
(d) four-jet inclusive events. Histograms show the {\sc pythia}
simulation normalized (increased by a factor of 1.3) to the inclusive two-jet 
sample at {\Et} $>$ 40 GeV.
}
\label{fig:et}
\end{figure}

\begin{figure}[h]
\vspace {-3.0cm}
\mbox{\hspace*{1.0cm}\epsfig{figure=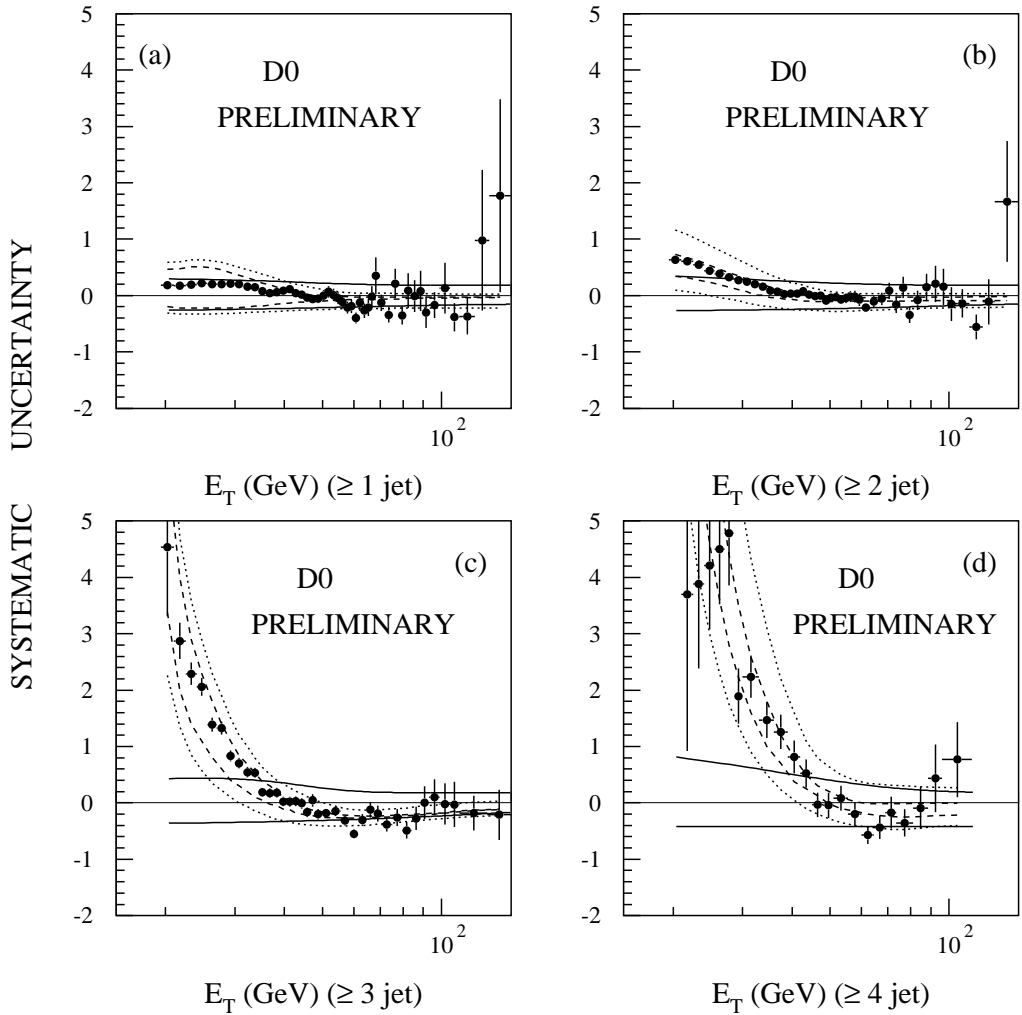 
,width=15cm}}
\caption{
(Data - Theory)/Theory as a function of the transverse energy of the 
leading jet for (a) single-jet inclusive, (b) two-jet inclusive, 
(c) three-jet inclusive and (d) four-jet inclusive event samples. }
\label{fig:et_prim_sys}
\end{figure}

Distributions in transverse energy for the leading 
jet for n=1 to n=4 inclusive jet events are shown in Fig.~\ref{fig:et}, 
together with the results from {\sc pythia} simulations. In these 
and all other plots, the data has been corrected for inefficiencies and 
energy calibration, but not for contributions from an underlying event. 
Also, we normalize the theory (increased by a factor of 1.3) to 
the observed two-jet cross section Fig.~\ref{fig:et}(b) for 
${E_{\rm T}} > 40 \rm ~GeV$. Figure~\ref{fig:et_prim_sys} shows the 
fractional difference (Data - Theory) / Theory for the {\Et} spectra in 
Fig.~\ref{fig:et} with the systematic uncertainties, arrising from 
uncertainties in jet-energy calibration and resolution. The theory is 
in agreement with the data for the single-inclusive jet sample in the 
entire {\Et} interval, and with the two-jet sample for most of the 
energy interval (some excess of data is observed at low {\Et}). However, 
for the three and four-jet samples, there is large excess relative to 
theory at low {\Et} and a deficit near 75 GeV. The shapes of the 
experimental and theoretical spectra are clearly different, and not 
reconcilable through re-normalization. 

The systematic uncertainty on the cross section is due primarily to the 
uncertainty in the energy calibration. The uncertainty from energy 
resolution represents the main uncertainty in the MC. The uncertainty 
from the energy calibration can be estimated by considering spectra 
with $\pm 1$ standard-deviation corrections to jet $E_{\rm T}$. The 
same procedure can be used to derive the uncertainty due 
to jet resolution in the MC. At 25 GeV uncertainty in the 
three-jet cross section due to calibration is 36\%, and the uncertainty 
in the MC due to resolution is 17\%. In Fig.~\ref{fig:et_prim_sys}, the 
relative systematic uncertainties corresponding to the energy 
calibration added to a 15\% uncertainty in luminosity are shown (in 
quadrature) by the solid lines centered about zero. The uncertainties 
from  energy smearing are shown by the dashed lines near the data 
points. The total systematic uncertainties on 
the ratio are shown by the dotted lines. Because the systematic 
uncertainties are highly correlated in {\Et} (a change of the cross 
section in one bin is accompanied by a corresponding change in 
neighboring bins), the departure of the ratio from zero cannot be 
explained solely by systematic uncertainties.  

\begin{figure}[h]
\mbox{\hspace*{1.0cm}\epsfig{figure=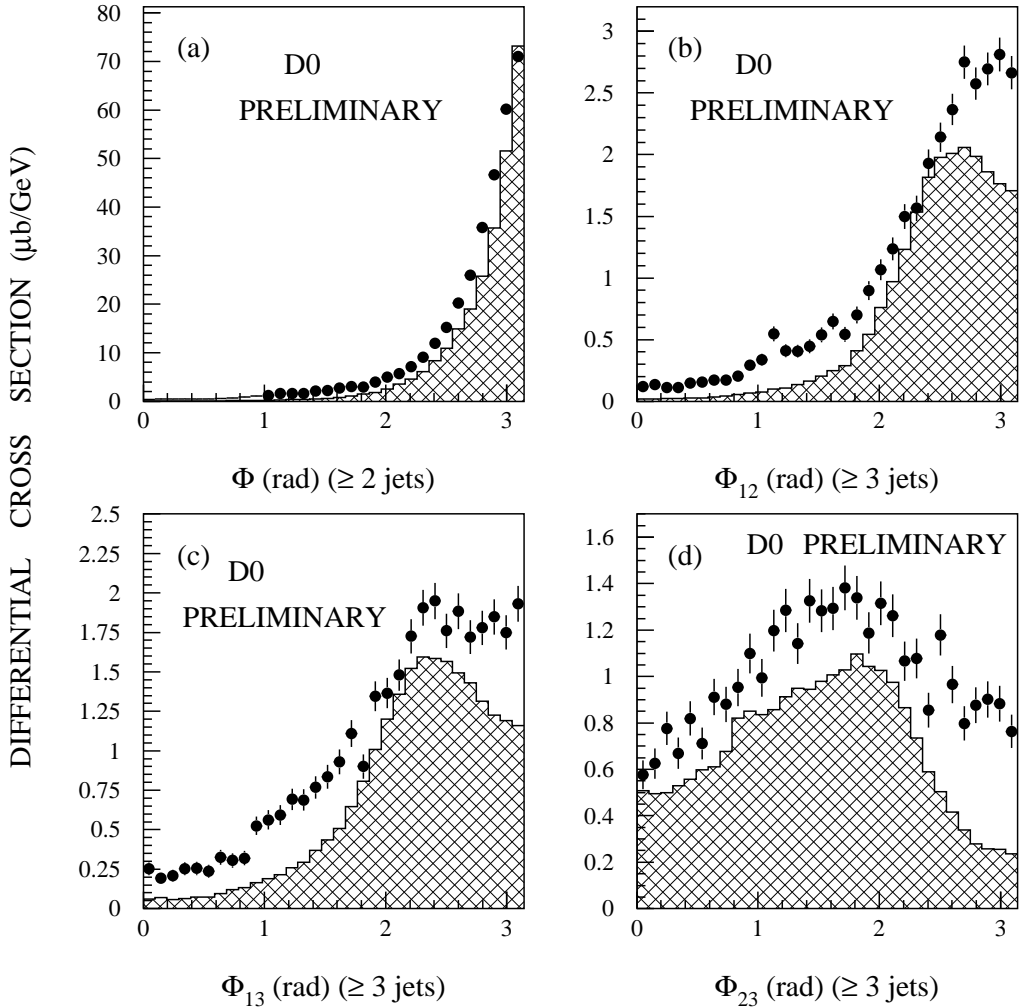,width=15cm}}
\caption{
Distributions of the relative azimuthal angle between two jets in 
(a) two-jet events and in three-jet events (b-d). Jets are ordered by 
their transverse energies. Histograms are from a {\sc pythia} 
simulation of such events. 
}
\label{fig:azim}
\end{figure}

To explore the discrepancies in three and four-jet 
production, we turn to observations of azimuthal distributions, 
distributions in summed transverse momenta, and three-jet studies. 
In Fig.~\ref{fig:azim}(a) we plot the azimuthal difference between the 
leading two jets in events with two or more jets. 
Figures~\ref{fig:azim}(b)-\ref{fig:azim}(d) show the azimuthal 
difference between the first and second, first and third, and second 
and third highest-{\Et} jets in a three-jet event. In 
Fig.~\ref{fig:azim}(a) we see the strong anticorrelation (in the 
transverse plane) expected of two-jet events. The distribution widens 
substantially in the three-jet sample 
(Fig.~\ref{fig:azim}(b)-\ref{fig:azim}(d)). The peaks correspond to the 
kinematic constraint of transverse momentum conservation for jets 
produced in hard QCD subprocesses. Altough, in general, {\sc pythia} 
reproduces the observed shapes, there is a large excess of events 
in the three-jet sample not consistent with expectation. In particular, 
there is a significant contribution to three-jet events with two jets 
back-to-back in the transverse plane near ($ \Phi = \pi $).

\begin{figure}[h]
\vspace {-2.0cm}
\mbox{\hspace*{1.0cm}\epsfig{figure=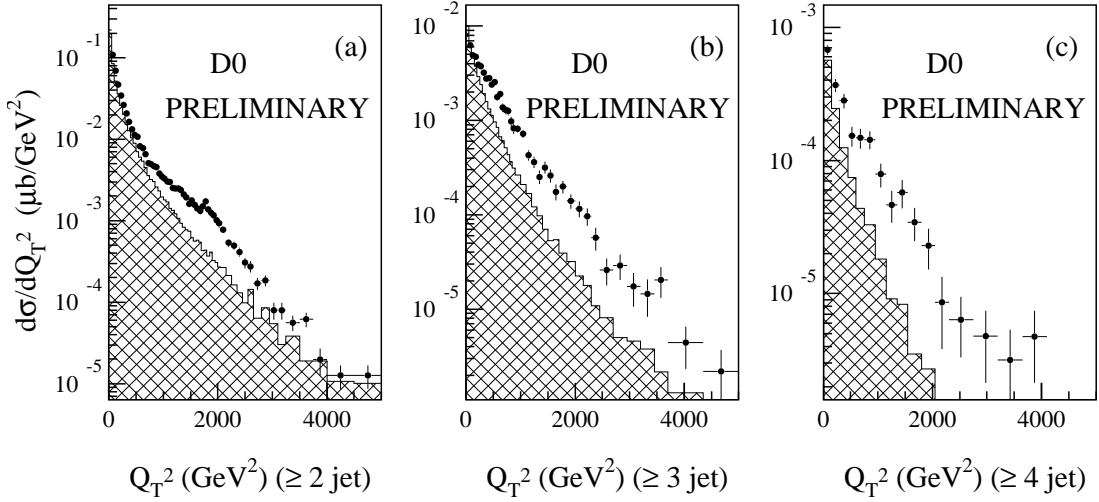,width=15cm}}
\caption{
Distributions in the square of the summed vector transverse momenta 
$Q_{T}^2$, for two-jet inclusive, three-jet inclusive and 
four-jet inclusive event samples (a-c). Histograms show the 
{\sc pythia} simulation.
}
\label{fig:q_T}
\end{figure}

Distributions in the square of the summed vector transverse momenta of 
jets $Q_{T}^2=({\bf E}_{T1}+{\bf E}_{T2}
+\cdots +{\bf E}_{Tn})^2$ shown in Fig.~\ref{fig:q_T}(a-c) indicate 
that the excess corresponds to events with a large imbalance. In fact, 
when these events are removed (with a requirement of a good balance 
in transverse momentum), the three-jet data and theory come into better 
agreement at small {\Et}. The shoulder at 
$Q_{T}^2 \sim 1600~\rm GeV^2$ in Fig.~\ref{fig:q_T}(a) is 
eliminated when the event sample is restricted to just two jets with 
{\Et} above 20 GeV, and no other jets between 8 and 20 GeV. This 
shoulder can consequently be associated with higher-order radiation. 
 
\newpage
\vspace*{1.0cm}

\begin{figure}[h]
\vspace {-3.0cm}
\mbox{\hspace*{1.0cm}\epsfig{figure=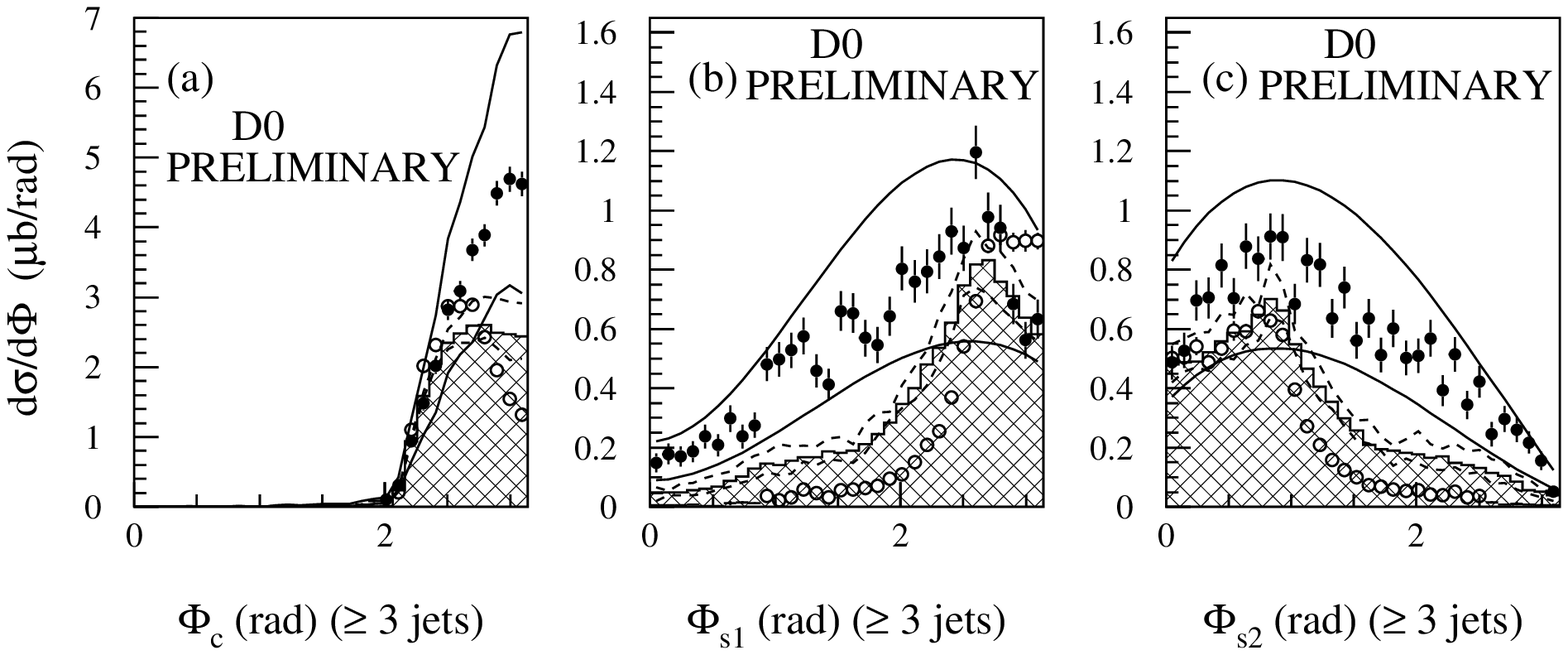
,width=15cm}}
\caption{
The distribution of the relative azimuthal angle in three-jet events, 
between the jets in the pair with the minimal scaled summed transverse 
momentum (a), and between the third jet 
and the other two leading {\Et} jets in the pair (b-c). Histograms show 
the {\sc pythia} simulation, and the open symbols the {\sc jetrad} 
simulation. 
}
\label{fig:phic_phi}
\end{figure}

To find the pair of jets $\{i,j\}$ most likely to originate from the 
hard interaction (rather then from gluon Brehmsshtrahlung), we define the 
scaled summed dijet vector transverse momenta: 
${\bf q}_{ij}=({\bf E}_{Ti}+{\bf E}_{Tj})/(E_{Ti}+E_{Tj})$. 
We choose the pair with the smallest magnitude of this vector 
and plot the distribution of the relative azimuthal angle $ \Phi _c $ 
between the jets in that pair Fig.~\ref{fig:phic_phi}(a). The data 
lie above theory in the region where two jets, reflecting a hard scatter, 
appears back-to-back ($\Phi_c = \pi $). {\sc pythia} shows a broader 
distribution, and the prediction from {\sc jetrad} is peaked away from 
$\Phi_c = \pi $ due to the presence of the third (radiated) jet.
\newpage
\vspace*{1.0cm}

Figures~\ref{fig:phic_phi}(b) and Fig.~\ref{fig:phic_phi}(c) show the 
azimuthal separation of the third jet from each of the two jets that  
correspond to the minimum $q_{ij}^2$. These distributions contain 
events only for $\pi -\Phi _c \le 0.4$; that is, events in which the 
balanced jets are essentially back-to-back. When the third jet is 
correlated with the balanced jets, it will be expected to be emitted 
along or opposite to the balanced jets. The uncertainties from the 
energy calibration and luminosity are shown by the solid lines, and 
from the energy resolution by dashed lines. We see that the data has 
a wider distribution than {\sc pythia}, and much wider distribution 
than {\sc jetrad}. The third jet appeares to be uncorrelated with the 
balanced jets, and is emitted at all angles. Our studies indicate that 
the observed differences in shape are not sensitive to modeling of the 
underlying event or contributions from multiple-parton scattering.  
\newpage

\vspace*{-2.0cm}

In summary, our data on multiple-jet production at low {\Et} show 
significant discrepancies with {\sc pythia} and {\sc jetrad}. This is 
observed in the distributions of the transverse energy of the leading 
jets (Fig.~\ref{fig:et}), in the square of the summed vector transverse 
momenta $Q_{T}^2$ (Fig.~\ref{fig:q_T}), and in the three-jet 
angular distributions that suggest the presence of an uncorrelated jet 
(Fig.~\ref{fig:phic_phi}). Additional corrections to QCD calculations 
are therefore required to accommodate these results; higher-order or 
{\sc BFKL} processes are possible candidates.  
%
%

We thank the staffs at Fermilab and collaborating institutions for
their contributions to this work, and acknowledge support from the
Department of Energy and National Science Foundation (U.S.A.),
Commissariat  \` a L'Energie Atomique (France), State Committee for
Science and Technology and Ministry for Atomic    Energy (Russia),
CAPES and CNPq (Brazil), Departments of Atomic Energy and Science and
Education (India), Colciencias (Colombia), CONACyT (Mexico), Ministry
of Education and KOSEF (Korea), and CONICET and UBACyT (Argentina). 
%

\newpage
\vspace*{1.0cm}
\vspace*{2.0cm}

\end{document}

%% file: list_of_authors_1_june_2001.tex
%
V.M.~Abazov,$^{23}$                                                           
B.~Abbott,$^{58}$                                                             
A.~Abdesselam,$^{11}$                                                         
M.~Abolins,$^{51}$                                                            
V.~Abramov,$^{26}$                                                            
B.S.~Acharya,$^{17}$                                                          
D.L.~Adams,$^{60}$                                                            
M.~Adams,$^{38}$                                                              
S.N.~Ahmed,$^{21}$                                                            
G.D.~Alexeev,$^{23}$
% ...
\\                                                                            
\vskip 0.30cm                                                                 
\centerline{(D\O\ Collaboration)}                                             
\vskip 0.30cm                                                                 
\centerline{$^{1}$Universidad de Buenos Aires, Buenos Aires, Argentina}       
\centerline{$^{2}$LAFEX, Centro Brasileiro de Pesquisas F{\'\i}sicas,         
                  Rio de Janeiro, Brazil}                                     
\centerline{$^{3}$Universidade do Estado do Rio de Janeiro,                   
                  Rio de Janeiro, Brazil}                                     
\centerline{$^{4}$Institute of High Energy Physics, Beijing,                  
                  People's Republic of China}                                 
\centerline{$^{5}$Universidad de los Andes, Bogot\'{a}, Colombia}             
\centerline{$^{6}$Charles University, Center for Particle Physics,            
                  Prague, Czech Republic}                                     
\centerline{$^{7}$Institute of Physics, Academy of Sciences, Center           
                  for Particle Physics, Prague, Czech Republic}               
\centerline{$^{8}$Universidad San Francisco de Quito, Quito, Ecuador}         
\centerline{$^{9}$Institut des Sciences Nucl\'eaires, IN2P3-CNRS,             
                  Universite de Grenoble 1, Grenoble, France}                 
\centerline{$^{10}$CPPM, IN2P3-CNRS, Universit\'e de la M\'editerran\'ee,     
                  Marseille, France}                                          
\centerline{$^{11}$Laboratoire de l'Acc\'el\'erateur Lin\'eaire,              
                  IN2P3-CNRS, Orsay, France}                                  
\centerline{$^{12}$LPNHE, Universit\'es Paris VI and VII, IN2P3-CNRS,         
                  Paris, France}                                              
\centerline{$^{13}$DAPNIA/Service de Physique des Particules, CEA, Saclay,    
                  France}                                                     
\centerline{$^{14}$Universit{\"a}t Mainz, Institut f{\"u}r Physik,            
                  Mainz, Germany}                                             
\centerline{$^{15}$Panjab University, Chandigarh, India}                      
\centerline{$^{16}$Delhi University, Delhi, India}                            
\centerline{$^{17}$Tata Institute of Fundamental Research, Mumbai, India}     
\centerline{$^{18}$Seoul National University, Seoul, Korea}                   
\centerline{$^{19}$CINVESTAV, Mexico City, Mexico}                            
\centerline{$^{20}$FOM-Institute NIKHEF and University of                     
                  Amsterdam/NIKHEF, Amsterdam, The Netherlands}               
\centerline{$^{21}$University of Nijmegen/NIKHEF, Nijmegen, The               
                  Netherlands}                                                
\centerline{$^{22}$Institute of Nuclear Physics, Krak\'ow, Poland}            
\centerline{$^{23}$Joint Institute for Nuclear Research, Dubna, Russia}       
\centerline{$^{24}$Institute for Theoretical and Experimental Physics,        
                   Moscow, Russia}                                            
\centerline{$^{25}$Moscow State University, Moscow, Russia}                   
\centerline{$^{26}$Institute for High Energy Physics, Protvino, Russia}       
\centerline{$^{27}$Lancaster University, Lancaster, United Kingdom}           
\centerline{$^{28}$Imperial College, London, United Kingdom}                  
\centerline{$^{29}$University of Arizona, Tucson, Arizona 85721}              
\centerline{$^{30}$Lawrence Berkeley National Laboratory and University of    
                  California, Berkeley, California 94720}                     
\centerline{$^{31}$University of California, Davis, California 95616}         
\centerline{$^{32}$California State University, Fresno, California 93740}     
\centerline{$^{33}$University of California, Irvine, California 92697}        
\centerline{$^{34}$University of California, Riverside, California 92521}     
\centerline{$^{35}$Florida State University, Tallahassee, Florida 32306}      
\centerline{$^{36}$University of Hawaii, Honolulu, Hawaii 96822}              
\centerline{$^{37}$Fermi National Accelerator Laboratory, Batavia,            
                   Illinois 60510}                                            
\centerline{$^{38}$University of Illinois at Chicago, Chicago,                
                   Illinois 60607}                                            
\centerline{$^{39}$Northern Illinois University, DeKalb, Illinois 60115}      
\centerline{$^{40}$Northwestern University, Evanston, Illinois 60208}         
\centerline{$^{41}$Indiana University, Bloomington, Indiana 47405}            
\centerline{$^{42}$University of Notre Dame, Notre Dame, Indiana 46556}       
\centerline{$^{43}$Iowa State University, Ames, Iowa 50011}                   
\centerline{$^{44}$University of Kansas, Lawrence, Kansas 66045}              
\centerline{$^{45}$Kansas State University, Manhattan, Kansas 66506}          
\centerline{$^{46}$Louisiana Tech University, Ruston, Louisiana 71272}        
\centerline{$^{47}$University of Maryland, College Park, Maryland 20742}      
\centerline{$^{48}$Boston University, Boston, Massachusetts 02215}            
\centerline{$^{49}$Northeastern University, Boston, Massachusetts 02115}      
\centerline{$^{50}$University of Michigan, Ann Arbor, Michigan 48109}         
\centerline{$^{51}$Michigan State University, East Lansing, Michigan 48824}   
\centerline{$^{52}$University of Nebraska, Lincoln, Nebraska 68588}           
\centerline{$^{53}$Columbia University, New York, New York 10027}             
\centerline{$^{54}$University of Rochester, Rochester, New York 14627}        
\centerline{$^{55}$State University of New York, Stony Brook,                 
                   New York 11794}                                            
\centerline{$^{56}$Brookhaven National Laboratory, Upton, New York 11973}     
\centerline{$^{57}$Langston University, Langston, Oklahoma 73050}             
\centerline{$^{58}$University of Oklahoma, Norman, Oklahoma 73019}            
\centerline{$^{59}$Brown University, Providence, Rhode Island 02912}          
\centerline{$^{60}$University of Texas, Arlington, Texas 76019}               
\centerline{$^{61}$Texas A\&M University, College Station, Texas 77843}       
\centerline{$^{62}$Rice University, Houston, Texas 77005}                     
\centerline{$^{63}$University of Virginia, Charlottesville, Virginia 22901}   
\centerline{$^{64}$University of Washington, Seattle, Washington 98195}       
\centerline{$^{65}$Petersburg Nuclear Physics Institute, Gatchina, Russia}